\documentclass{kapproc} 

\usepackage{procps}

\usepackage[dvips]{graphicx}

\nochapequationnumber 

\let\footnote\savefootnote

\let\footnoterule\savefootnoterule

\normallatexbib 

\let\lcitebracket[
\let\rcitebracket]

\begin{document}

\articletitle{Photon echoes of molecular \\
photoassociation\footnote{Proceedings of  SPIE Vol.3734, 86-95 (1999)}}
\chaptitlerunninghead{Photon echoes of molecular photoassociation}
\rhead{Photon echoes of molecular photoassociation}
\author{Alexander G.Rudavets\footnote{address for correspondences:  Arudavets@mics.msu.su}
\altaffilmark{1}}
\altaffiltext{1}{Moscow Institute of Physics and
Technology, 141700 Dolgoprudny, Russia}
\author{Alexander M.Dykhne \altaffilmark{1,2}}
\altaffiltext{2}{TRINITI, 142092 Troitsk, Russia}
\begin{abstract}

Revivals of  optical coherence of  molecular photoassociation driven by two
ultrashort laser pulses are addressed in the Condon approach. Based on
textbook examples and numerical simulation of KrF excimer molecules, a
prediction is made about an existence of photon echo on free-bound
transitions. Delayed rise and fall of nonlinear polarization in the
half-collisions are to be resulted from the resonant quantum states
interference whether it be  in gas, liquid or solid phases.
\end{abstract}
\section{Introduction}
At present, the photoassociation spectroscopy (PAS) of atoms
colliding along the ground state potential and interacting with
photons to produce molecular dimers (through the inverse
predissociation) has  attracted a great deal of attentions
$^{1-7}$. This technique, as applied in conventional continuous
wave$^{5, 16}$ (CW) and fentosecond$^1$ regimes, permits a
determination of the rovibrational leveles, hyperfine structure,
scattering lengths and/or dynamics of bond formation that can be
derived with an unprecedented accuracy. To assess the primary
processes  of harpooning reactions$^2$, photoassociative and the
Penning ionizations $^3$, relaxation of the van der Waals
complexes$^4$ in liquids and quantum  gases in optical traps, the
PAS is ideally suited.  The review of earlier developments, which
have been collected under the rubric of chemical reactions in
transition spectroscopy, can be found in $^6$.  The most recent
innovation to the CW PAS is investigation of  Bose's state of the
cold atoms$^7$ in which a loss of coherence can be controlled
trough the laser assisted binary collisions.

One would naively claim that the femtosecond PAS is not quite
sensitive as the resonant CW schemes,  since only a low number of
atoms distributed randomly in close proximity during ultrashort
pulses may contribute to the coherent bonding. However,  the
shorter impulse, the wider its spectrum, the more the free-bound
transitions involved. An example is found in photoinduced
harpooning reactions $^8$ in which according to the M. Polanyi the
attacking open shell atom tosses out its valence electron, then
hooks the closed shell atom and hauls it in with the Coulomb force
even at a range of hundreds Angstroms\footnote{The long-range of
the reaction increasingly grows at low temperature }. This
reaction endowed with separation of degrees of freedom into
nuclear and electronic coordinates controlled by the femtosecond
optical pulses may have a high quantum yield.

It is straightforward to extend the femtosecond PAS to address the four wave
mixing geometry or its two or three wave realizations. This makes it
possible to investigate the nonlinear  optical effects those are no longer
observable in the weak fields. Such as a photon echo photoassociation which
is counterpart of the photon echo of photodissociation. A theoretical
description of the latter is only being reported by us $^{9,10}$ in context
of the wave packet engineering. It is evident now, that a well matched
spreading and squeezing  of the vibrational wave packets are called for the
desired optical properties of molecular samples. So that the quantum control
in the weak field regime requires a complicated pulse shaping $^{11}$. While
"spontaneous" coherence, i.e., a photon echo, is obtained trough a more
strong, short and  transform limited (non chirped) fields  that enables one
to automate the wave packet preparation on the free-bound transitions.

The transients  having the entanglement of discrete levels and continuum
usually result in a decay $^{12}$. The quantum transition amplitudes may
decrease depending on the contribution of the molecular repulsion or quantum
dispersion. In the frequency domain, the resonant coupling is accompanied by
the inhomogeneous broadening (dephasing) of the Franck-Condon transitions.
If the latter dominates over an irreversible relaxation in multi-channel
(chemical) collisions, spontaneous emission, the photon echo can be
harvested as for the bound-bound transitions $^{13}$.

To make this point a bit clear, we follow the next plan. In Sec.2 the Condon
model of femtosecond PAS is analyzed. We relate the delayed transients to
the wave packet interference residing in molecular continuum after the
second impulse. Analytical expressions are derived for the $\chi^{(3)}$
polarization induced by 2 $\delta$-like impulses. A singularity of
$\chi^{(3)}$  is identified with an infinite region of the Franck-Condon
transitions simultaneously involved. For the realistic field shapes this
trait is eliminated. The PAS is particular appealing for the investigating
excitations of the rare gas halides, for which the electronic ground states
are dissociating. This system is of use as an active media of the high-power
lasers\footnote{Europe's largest KrF excimer laser "TITANIA"
(http://www.clf.rl.ac.uk/index.html) has an output of 10$^{19}$W/cm$^2$ and
pulses duration 350 $fs$ } and photon echo of photoassociation might deliver
a valuable information on timescales of excited molecular states in the
lasing. Specifically, the relevant model of molecular dynamics of the
krypton fluoride KrF(B) excimer is simulated in Sec.3, where we consider a
photoinduced harpooning reaction of interacting Kr cations and F anions
created at thermal conditions.  To avoid the Condon reflection principle,
which overestimates contribution of turning points and disregards
oscillatory tails of the wave functions, the B-X transition (at 248 nm)  is
treated in the globally uniform approach. The Wigner representation gives an
insight into why the photon echoes of photoassociation exist from the
kinematical point of view. Section 4 concludes.
\section{Condon Model}
Strictly speaking, {\it 'ab-initio'} approach to PAS necessitates a
many-body field theory of interacting matter-light species. However, any
more rigorous treatment of the photoassisted reactions beyond the
Born-Oppengheimer strategy faces sophisticated mathematical problems of
exclusion of extra degrees of freedoms and transitions between
multidimensional surfaces. It is not surprising, then, that a generic model
with simplified potential curves, in which the bound-free transitions occur
in one-body treatment, has been introduced at the drawn of the age of
quantum mechanics $^{14}$. Herein, a separation of degrees of freedom gives
the paradigm of application of the Franck-Condon principle asserting that
for a time of electron transfer, the coordinates and momenta of nuclei do
not vary because of the large difference between the electronic and nuclear
masses. Molecular dynamics is governed by the Hamiltonian operators:
$$
\hat H_{b}=\hat T_{kin}+\hat V_{b}; \quad
\hat H_{f}=\hat T_{kin}+\hat V_{f},
$$
where $\hat T_{kin}$ is the operator of kinetic energy of relative
motion in the motionless center mass of a molecule and the
Doppler's shift of frequency is disregarded as marginal.
Initially, atoms collide along the ground electronic term with
constant interatomic interaction $\hat V_{f}=0$. The excited
electronic curve can be presented by the potential $\hat
V_{b}=\omega_0 + \Omega^{2} x^2/2$ with understanding that the
actual curves are deflected of the parabolic well at longer and
shorter distances from its equilibrium position. The model
potential has merit of being exactly solved best serving for our
methodological purpose. We have chosen the  test example of
oscillatory motion with frequency $\Omega$   as a vehicle with
which to demonstrate general aspects of photoassociation. Without
restricting the generality, an effective mass of molecule equals
unit. The electronic states are coupled by the electric dipole
interaction $\hat V_{bf}={\hat \mu } E(t)$, where the laser
impulses have the slow envelops $E(t)$ and fast optical frequency
$\omega_0$. The envelope differs from zero only for the moments of
photon impacts.

With the rotating wave approximation, the Schr\"{o}dinger equation for the
bound states  $\vert \psi_b\rangle$ and scattering states $\vert
\psi_f\rangle$  can be written in the notation of Dirac as
\begin{eqnarray}
{i\hbar}\vert\dot\psi_{b}\rangle & =&
\hat H_{b}\vert\psi_{b}\rangle +
\hat V_{bf}\vert\psi_{f}\rangle\nonumber\\
{i\hbar}\vert\dot\psi_{f}\rangle & = &
\hat H_{f}\vert\psi_{f}\rangle +
\hat V_{fb}\vert\psi_{b}\rangle .
\end{eqnarray}
In order to have general formalism capable of dealing with both
bound and scattering states,  we employ  the momentum
representation. A crucial assumption is made that our system has a
very large box volume $V$. The energy spectrum of atoms forming
the plane waves  is bound to be quasi-continuous in the box. It is
required of the initial quantum conditions $\langle
p\vert\psi_{f}(0)\rangle= \psi_{f}^{0}(p)$ that the box wave
functions possess the property of random phases. This requirement
corresponds to the "molecular chaos" with the canonical Gibs
thermal distribution in the atomic ensemble:
$$
\langle  \stackrel{*}{\psi_{f}^{0}}(p_{1})\psi_{f}^{0}(p)\rangle_{ave}
=Z^{-1}e^{-\beta H_{f}(p)/\hbar}\delta(p-p_1),
$$
where $\beta$ is the inverse temperature, Z is the statistical
sum. The box normalization (a molecule per the box volume) is
implied.  The quasi-continuous states correlate initially only
with the coincident momenta giving the Gibbs density, while the
bound molecular states are  empty $\vert \psi _{b}(0) \rangle =0$
before the laser excitation.

The resonant states between the laser impulses are governed by
the molecular evolution operators
\begin{eqnarray}
\vert\psi_{b}(t)\rangle & = & e^{-i t\hat{ H}_{b}/\hbar}
 \vert\psi_b(0)\rangle \nonumber\\ 
\vert\psi_{f}(t)\rangle & = & e^{- i t\hat
H_{f}/\hbar}\vert\psi_f(0)\rangle,
\end{eqnarray}
where the momentum representation can be used to arrive at the matrix elements
\begin{eqnarray}
{\langle} p \vert e^{-i\frac{t\hat{ H}_{f}}{\hbar}}\vert p_{1}\rangle & = &
e^{-i t p^{2}/2\hbar}\delta(p-p_{1}) \nonumber \\
\langle p \vert e^{-i\frac{t\hat{ H}_{b}}{\hbar}} \vert p_{1}\rangle & = &
\frac{e^{i ((p^{2}+p_{1}^{2})cos(\Omega t)-2pp_{1})/(2\hbar \Omega
sin(\Omega t))}}{\sqrt { 2i\hbar\Omega \pi sin(\Omega t)}}.
\end{eqnarray}
The laser field couples the resonant electronic states $\vert
\psi_b\rangle$ and $\vert \psi_f\rangle$ and to prepare their
mixing in the limit of a weak,  short impulse  according to eq.
(1)
\begin{eqnarray}
\vert \psi_b(\tau)\rangle &=&-i\theta\vert \psi_{f}^{0} \rangle
\nonumber\\
\vert\psi_f(\tau)\rangle &=&\vert \psi_{f}^{0}\rangle-i\theta\vert
\psi_{b}(\tau)\rangle,
\end{eqnarray}
where zero time $t=0$ corresponds to the beginning of an ultrashort
pulse of the
duration $\tau $, so that $\theta=\tau \mu \cdot E/\hbar \ll\pi$ is the
pulse area.
\medskip
\par
Let us consider an idealization of the $\delta$- like impulses.
The Franck-Condon principle ensures the intact positions and
momenta of the wave packets on the resonant electronic terms
during the laser field action. The optical polarization of the
system is found by calculating the dipole moment $\mu$ in the
Condon approximation in which a total span of Franck-Condon region
of free-bound transition includes the (infinite) box V, whilst the
relative electronic transition moment, i.e., the dipole operator
$\hat\mu$ is independent of a position of atoms in the box:
\begin{equation}
\langle \hat \mu(t)\rangle_{ave}=\mu \langle\langle
\psi_{f}(t)\vert \psi_{b}(t)\rangle\rangle_{ave}.
\end{equation}
The free decay of the optical polarization after the first laser impulse
follows the time dependent Fermi golden rule:
\begin{equation} \langle
\hat \mu^{(1)}\rangle_{ave}=-i\mu\theta Tr[e^{-i t\hat
H_{b}/\hbar}e^{i(t+i\beta)\hat H_f/\hbar}]/Z =-i\mu\theta K^{(1)}(t),
\end{equation}
where the indentity relation $ \hat I=\int dp \vert p \rangle \langle p
\vert $ and the matrix elements of Eq. (3) are applied for the trace
calculation
\begin{eqnarray}
K^{(1)}(t) &=&Z^{-1}[2\pi\hbar\Omega sin(\Omega t) ]^{-1/2}\int dp
e^{i[i\beta+t+tg(\Omega t/2)/\Omega]p^{2}/2\hbar}=\nonumber\\&=&
Z^{-1}[cos(W_{decay}(t))-1]^{-1/2}.
\end{eqnarray}
The kernel  $K^{(1)}$ is determined by a "hyperbolic" rotation on
complex angle $W(t)$ given by:
\begin{equation}
cos(W_{decay}(t))=cos(\Omega t)+
0.5\Omega(i\beta+ t)sin(\Omega t),
\end{equation}
The hyperbolic rotation is inspired by the dynamical symmetry group
$SU(1,1)$ responsible for the phase and amplitude modulation of average
dipole moment.  In frequency domain, the hyperbolic rotation gives rise  in
inhomogeneous broadening of spectral lines of free-bound transitions
$^{15}$.

The free-bound transition amplitude $K^{(1)}(t)$ has a time-integrable
singularity of the order $1/\sqrt{t}$. Notice that this kernel is resulted
from an unbounded spectrum of the simultaneously driven resonant transitions
induced by the $\delta$-pulse. This artifact of the polarization singularity
can be corrected whether by a dipole moment cutoff that leads us beyond the
Condon approximation and/or frequency limited spectrum of the excitation
field. However, being the $\mu$ cutoff length dependent, it becomes tedious
for the analytical analysis. The thermalization and internal molecular
interference add to the complexity of the problem. This is the case of
femtosecond PAS. The nonlinear optical polarization depends also on the
multi-dimensional convolutions of the actual field envelopes with the
singular kernels induced by the $\delta$ fields.  Their nonlinear
combinations provide another vectors of directional coherence along which
the echo signals are to be observed. We shall set these problems aside for
sake of simplicity to treat molecular photoassociation in the $\delta$-pulse
limit and focusing on collinear geometry of incident electromagnetic wave
vectors. Suggesting that these two $\delta$-pulses operate having the
inter-pulse delay $T$, we find the mixing bound and free states as
\begin{eqnarray}
\vert \psi_f(T)\rangle &=&[e^{-i T\hat H_{f}/\hbar} -\theta_1\theta_2 e^{-i
T\hat H_{b}/\hbar}]\vert \psi_{f}^{0}\rangle, \nonumber\\ \vert
\psi_b(T)\rangle &=&-i\theta_2 e^{-i T\hat H_f/\hbar}\vert
\psi_{f}^{0}\rangle,
\end{eqnarray}
where the initial state $\vert \psi_{f}^{0}\rangle$ belongs to the Gibbs
distribution of collision pairs. By  $\theta_1$ and $\theta_2$ we denote the
areas of the first and the second laser impulses respectively.  These pulses
govern a quantum superposition of the states
\begin{eqnarray}
\vert\psi_{f}(t)\rangle & = & [e^{- i t \hat H_f/\hbar}-\theta_1\theta_2
e^{-i(t-T)\hat H_f/\hbar} e^{- i T \hat
H_b/\hbar}]\vert\psi_{f}^{0}\rangle\nonumber\\&=&
\vert\psi_{f}^{0}(t)\rangle-\theta_1\theta_2\vert\bar
\psi_{f}^{0}(t)\rangle, \nonumber\\
\vert  \psi_b(t)\rangle &=& -i\theta_2 e^{- i(t-T) \hat H_b/\hbar} e^{-
iT\hat H_f/\hbar}\vert\psi_{f}^{0} \rangle  \nonumber\\&=&
-i\theta_2\vert\bar\psi_b(t)\rangle,
\end{eqnarray}
in which  the normalized state vectors $\vert \bar\psi_b(t)\rangle, \vert
\bar\psi_f(t)\rangle$ entering Eq.(10) with the coefficients proportional
to $\theta_2$ and $\theta_1 \theta_2$ yield a dominant contribution to the
delayed transition amplitudes.  We call these vectors as the "propagating"
states $^{10}$. The free-bound transients are presented by a series of the
propagation operators each of them being responsible for an adiabatic
step along the molecular curves. Therefore, a first nonlinear correction
to the dipole moment averaged over the Gibbs distribution can be cast as
\begin{eqnarray}
\langle \hat \mu^{(3)}\rangle_{ave} &=&i\mu\theta_{1}\theta_{2}^{2}
K^{(3)}(t),\nonumber\\
K^{(3)}(t) &=& Tr[e^{i\frac{(T-t)\hat H_{b}}{\hbar}}e^{i\frac{(t-T)\hat
H_f}{\hbar}}e^{-i\frac{T\hat H_{b}}{\hbar}} e^{-\frac{(\beta-iT)\hat
H_f}{\hbar}}]/Z.
\end{eqnarray}
Using the momentum representation of the propagation operators and
performing this 2-d Gauss integration over momenta in the matrix elements
Eqs.(3,13), we obtain a characteristic function
\begin{equation}
K^{(3)}(t) =Z^{-1}[cos(W_{echo}(t))-1]^{-1/2},
\end{equation}
in which a hyperbolic rotation owing to the 2-pulse interaction  is
\begin{eqnarray}
cos(W_{echo}(t))&=& cos(\Omega\bar t)+0.5\Omega((\bar t+i\beta)
sin(\Omega\bar t) \nonumber \\&+& \Omega(\bar t+T)\,(T
-i\beta)sin(\Omega(\bar t+T))\,sin(\Omega T)),
\end{eqnarray}
being conveniently expressed trough the shifted time $\bar t=t-2T$.
\medskip
\par
For the weak field regime the resonant coupling of continuum and bound wave
functions in the Franck-Condon region is determined by the time dependent
Fermi golden rule as shown in Fig. 1. A monotonic decay of the Franck-Condon
factors will no longer be valid at a low temperature, e.g. at inverse
temperature $\beta=1/\Omega$ corresponding to one vibrational quantum. The
reason has to do with the quantum dispersion which alone cannot destroy
coherence. Molecular repulsion for inclined potential curves or
predissociation would afford exponential decay $^{12}$. But even in this
instance, a net constructive or destructive interference of two wave packets
prepared by the pump and delayed impulses occurring at or near multiples of
the vibrational period might result in the revivals of coherence. The
transition amplitude spikes reflect the return of the vibrational wave
packet on the attractive potential curve.

To consider the echo effect  in the weak field regime   we derived  the
$\mu^{(3)}$- polarization from Eqs.(11-13). In the Condon model, the
transition amplitudes between the propagating states are obtained explicitly
as a function of time and an average polarization is plotted in inset in
Fig.1.
\begin{figure}
{\includegraphics[width=3.0in]{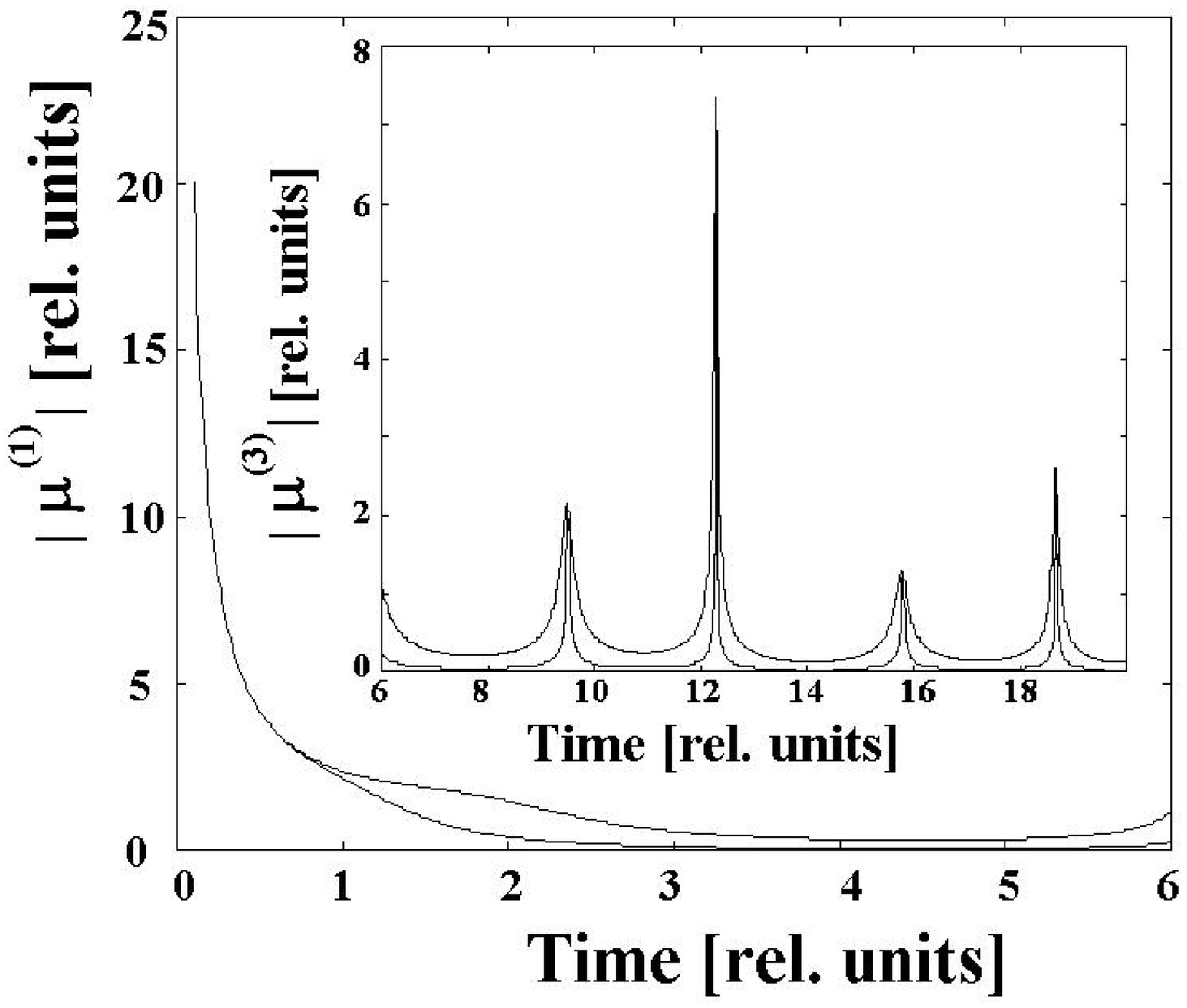}} \narrowcaption {The free radiation
decay and echo transients in the Condon model. The higher temperature, the
faster decay and the sharper the photon echo peaks. This behaviour is
illustrated by two curves at inverse temperature $\beta=1/\Omega $ and
$\beta=1/5\Omega $. Inset displays the $\mu^{(3)}$ dipole moment arising on
free-bound transition following the second ultrashot impulse which fires
with delay $T\approx 2\pi/\Omega $ matching molecular period. Use of
normalized impulse areas $\theta_{1,2}=1 $ and frequency $\Omega=1$ is made
here.}
\end{figure}
The echo signal itself consists of a number of short peaks most prominent
when a delay $T$ is close to multiple of the vibrational period ($T\approx
T_{vib}=2\pi/\Omega $). The Eq.(13) determines the positions of the
"spontaneous" revivals near $t=nT, (n+1/2)T \quad n=1,2...$. In the occasion
$\vert sin (\Omega T)\vert\ll 1$ and $ (\Omega T sin(\Omega T))^{2} > 1$, we
easily see that the half width at a half maximum (HWHM) at $t=2T$ is
estimated from Eq.(13) as
\begin{equation}
\tau_{echo}\approx \beta/\Omega T.
\end{equation}
This simple expression can be understood as a time taken by a classical
particle to move the thermal de-Broglie wavelength $r_{0}\propto
\sqrt{\beta}$ with velocity  $v=\Omega\langle R\rangle_{ave}$ depending
directly on the correlation radius $\langle R\rangle_{ave}$ of the
propagating state in the continuum of states,
$$ \langle
R(2T)\rangle_{ave}=\langle\langle\bar\psi_f(2T)\vert \hat R \vert
\bar\psi_f(2T)\rangle\rangle_{ave} \propto T v_{0},
$$
where $v_0 \propto 1/\sqrt{\beta}$ is the thermal velocity of atoms. So
that, we obtain the correlation time $ \tau_{cor}=r_{0}/Tv_{0}=\beta/\Omega
T,$ which matches Eq.(14) for HWHM  and has to play a key role in
determining the nonlinear polarization peaks. What this means is the
validity of kinematical argumentation  based on the quasiclassical transport
theory allowing us to justify the echo of photoassociation  just as the
photon echo of molecular photodissociation$^{10}$. It is appropriate to
recall that in the latter, the ground molecular wave function forms a wave
packet of the receding atoms following first impulse. Atomic distribution
acquires a stretched elliptic shape in the phase space formed by the
coordinates and velocities. The ellipse is strongly squeezed in the
transversal direction by virtue of "free" propagation. However, the total
phase space volume (according to the Liouville theorem) persists. Being
driven back by the second impulse, the wave packet replica settles in the
binding term as a non stationary state.

For the harmonic model, the ellipse rotates as a hole at the vibrational frequency. In
addition, the replica of the initial state, which was preserved after the first
impulse, repeats the evolution of the wave packet in the molecular continuum. The
transition amplitude $\vert K^{(3)}\vert ^2$ strongly depends on the interference
between the quantum-mechanical paths of the  wave packets driven in tandem. The last
manifests in squeezing of the photon echo peaks inversely quadratic of the delay $T$:
\begin{equation}
\tau_{echo}=
\tau_{decay}/((\Omega T)^2+1),
\end{equation}
where $\tau_{decay}$ is the time of free radiation decay depending
on molecular curves and temperature. It equals a molecular radius
$a$ divided by typical velocity of receding atoms in the phase
space. The longer the inter-pulse delay $T$, the larger the
squeezing of molecular states, the larger the correlation radius
of receding atoms, and the faster decay of the Franck-Condon
overlapping. This implies, that the wave packet moving on the
smaller ellipse diameter $a$ varying as $1/\Omega T$ has to move
it with velocity $v$ which grows directly $\Omega T$ (proportional
to the correlation radius). The overlap time is given by the ratio
$a/v$ to yield the mentioned trend $\tau_{echo}\propto (\Omega
T)^{-2}$.

The same reason holds true for the photon echo of photoassociation. But its duration
features its inverse power of $T$. This is because only a propagating state in the
continuum is being squeezed after the second impulse as a function of inter-pulse
delay. The difference between the echo scaling laws for the photodissociation and
photoassociation is founded on the fact, that, in the former, both wave packets
evolved on the binding term and in the continuum X-states are squeezed. Also  the
$\mu^{(3)}$- polarization has a maximum peak near $t=2T$. Our analysis indicates that
the HWHM increasingly decreases with the temperature. Besides, at the delay time
exactly matching the oscillatory period $2\pi/\Omega$, the amplitude of spontaneous
revivals may appear to diverge at points $t=2T$. Fortunately, this divergence rules
out in the $\mu^{(3)}$ polarization controlled by the realistic laser field which
shape differs from the $\delta$-like envelopes.

The theory of this phenomenon is best clarified in the phase space of the system. The
free-bound transitions are characterized by the distribution of the collisional pairs
over their electronic excitations, internuclear separations and velocities. Therefore
an uniform representation for the scattering states, those are quasiclassical in their
nature, may be helpful. Just as for the photodissociation picture, the molecular
dynamics can be visualized for the Wigner distribution functions.  For the thermal
atoms, the Wigner map has the shape of infinite strip along the coordinate axis. The
strip width in the transversal direction is equal to the thermal de-Broglie
wavelength.

The first impulse conveys the collisional pair to the binding electronic B-term. The
replica of the thermal Wigner distribution  sets here.  After that, the atoms
oscillate between the turning points. This corresponds to the strip distribution in
phase space which rotates at the harmonic frequency. The second impulse does the same
with the  remaining portion of colliding atoms. In addition, it simultaneously
promotes the rotated strip to the continuum, where the latter starts to spread along
the coordinate axis. The spreading distribution squeezes in transversal direction. It
is apparent, that their overlap sharply rises and falls, when the adjacent Wigner maps
cross each other. If the maps turn out parallel, a total span of the induced
free-bound transitions includes an infinite volume of our simplistic model enabling
the Franck-Condon factor to diverge.
\section{Wigner Functions of Photoassociation of Kr-F
Collisional Pairs } Consider the product of the density matrices
of the ground and excited electronic states. The time-delayed
nonvanishing overlap of the states in phase space would signature
an existence of photon echo of photoassociation. For ultrashort
pulses, the propagating states contribute to the delayed
transients.  Then, the density matrices
$\hat\rho_{b}\,=\langle\vert\psi_{b}\rangle\langle\psi_{b}\vert\rangle_{ave}$
and
$\hat\rho_{f}\,=\langle\vert\psi_{f}\rangle\langle\psi_{f}\vert\rangle_{ave}$
are related to the Wigner distribution function
$W_{{\alpha}}(p,q)$ via the Fourier transform in the coordinate
space
\begin{eqnarray}
W_{\alpha }(p,q)=(2\pi)^{-1} \int dr \,e^{ ipr}\, \langle\langle
q-r/2\vert\psi_{\alpha}\rangle\,\langle\psi_{\alpha} \vert
q+r/2\rangle\rangle_{ave} ,~\alpha=b,f.
\end{eqnarray}
The overlap of the Wigner distribution functions in the phase-space is
expressed as
\begin{eqnarray}
Tr[\hat \rho_{b}\,\hat \rho_{f}] =\int\!\!\int\,dpdq
\,\, W_{b}(p,q)\,W_{f}(p,q).
\end{eqnarray}
In our simple model, the quasiclassical functions
$W_{\alpha}(p,q)$ follow the Liouville transport equations
\begin{eqnarray} \frac{\partial W_\alpha }{\partial t}
+p\,\frac{\partial W_\alpha }{\partial q} -\frac{\partial V_\alpha
}{\partial q} \,\frac{\partial W_\alpha}{\partial p}=0,\end{eqnarray} with
an accuracy of $\hbar^2$ independent of the model potentials under study.
This approach guarantees an uniform treatment of molecular dynamics even
though a major contribution to the overlap integral is accumulated near the
turning points, where the standard quasiclassical theory fails. We take an
advantage of  the Wigner representation to calculate the matrix density
overlap  of Kr(4p$^6 S_{0}$)-F(2p$^5$ $^2$P) photoassisted collisions along
X- and B-molecular terms shown in Fig.2.
\begin{figure}{\includegraphics[width=3.in,height=3.3in]{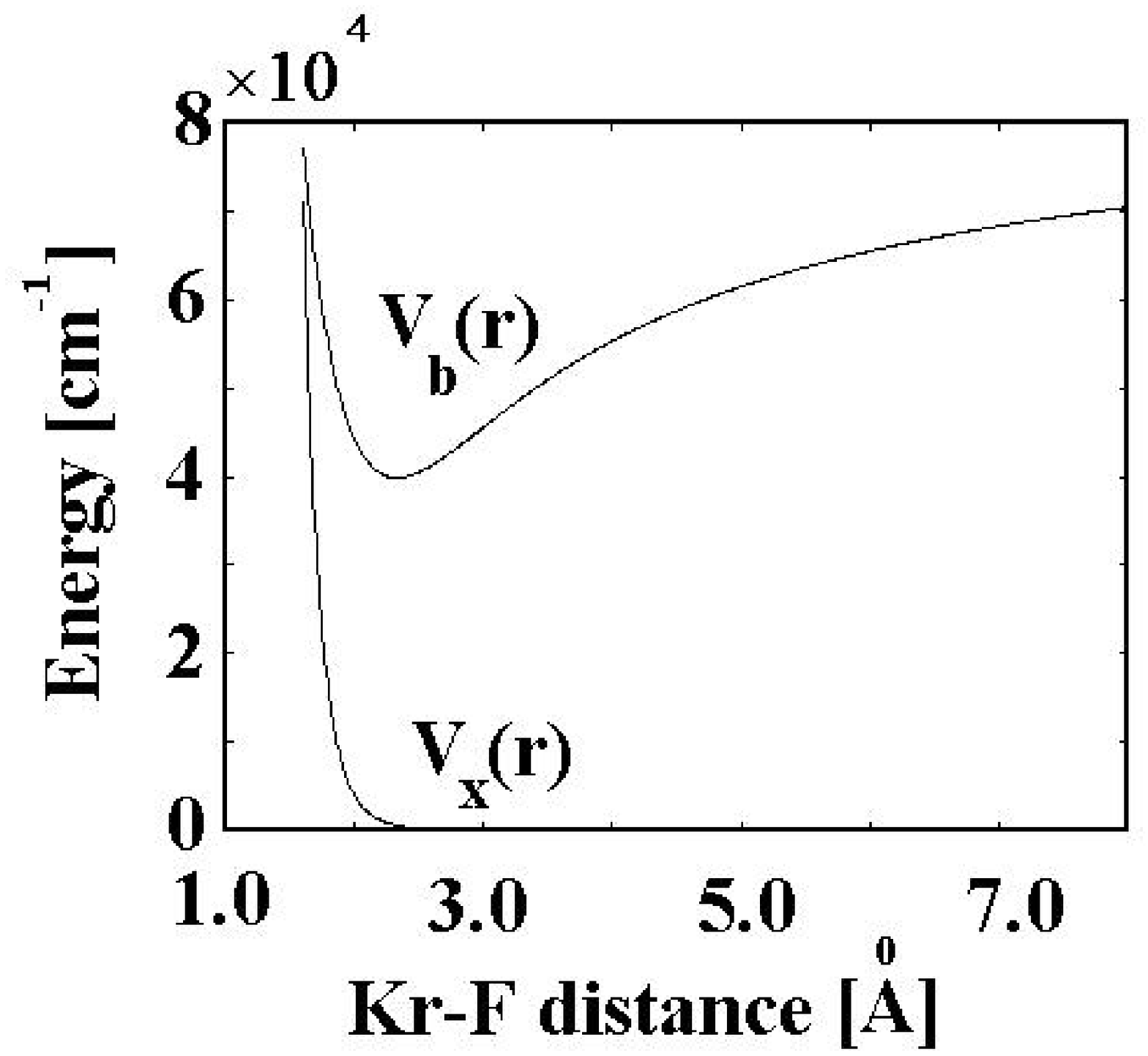}}
\narrowcaption{Interaction between Kr cation and F anion is described by the
truncated Ritner potential$^{16}$,
 $V_{b}(r)=a+b\,e^{-\gamma r}-c_{1}r^{-1}-c_{4}r^{-4},$ where r is the Kr-F distance, $a=85510
\,[cm^{-1}]$ is the separated ion limit, $\gamma=2.94\, [\AA^{-1}]$ and
$b=1.01\times10^{7}\, [cm^{-1}]$ are the constants of electron shells
exchange repulsion, $c_{1}=2.19\times 10^{5}\,[cm^{-1}\AA]$ and
$c_{4}=1.7\times 10^{5}\,[cm^{-1}\AA^4]$ are the parameters of Coulombic and
ion-quadrupole interactions respectively for the $B\Sigma_{1/2}^{+}$ state.
For the ground $X\Sigma_{1/2}^{+}$ state, the interaction potential is
assumed to be purely repulsive: $V_{x}(r)=c_{13}r^{-13}$, where
$c_{13}=3.2\times10^7 [cm^{-1}\AA^{13}]$.}
\end{figure}
Applying the classical dynamic routines to colliding Kr and F atoms, we
directly obtain the Wigner density functions of photoassociation. The
$\delta$-like envelopes of laser fields allow us to deal just with  the
propagating states density. In Fig. 3 we display the contour maps calculated
from the transport equation Eq. (17). The thermal atoms occupy the
half-strip terminating in a short distance due to the strong exchange
repulsion on the covalent X-term. The energy gap between the ground covalent
state and bottom of the ionic B-term is in range of 248 nm wavelength
resonant to the laser pulse frequency. The Wigner maps in the upper row in
Fig. 3 present the evolution of the thermal distribution of the X-states
distorted by the first $\delta$-pulse that resonantly drives them to ionic
B-states. Then, the F anions and Kr cations are attracted together due to
the joint action of the Coulomb force and induced ion-quadrupole
interactions. In the harpooning reaction, the Wigner map of the thermal
atoms  curves forming a hook and then a spiral structure in the phase space.
The spiral step is conditioned by the potential anharmonicity. For the pure
harmonic model, a rotation of the initial strip as a hole at the oscillatory
frequency takes place.
\begin{figure}
\caption{ The Wigner contour map of the density of KrF excimer molecules
immediately following an ultrashort pulse at 0 fs, 50 fs and 90 fs. Upper
panels show evolution of the Wigner functions due to a harpooning reaction
on the ionic B-curve. Lower panels show the thermal distribution of the
colliding atoms along the covalent X-curve. The contour lines are taken at
half maximum height of the Wigner's
functions}{\includegraphics[width=4.5in,height=3.5in]{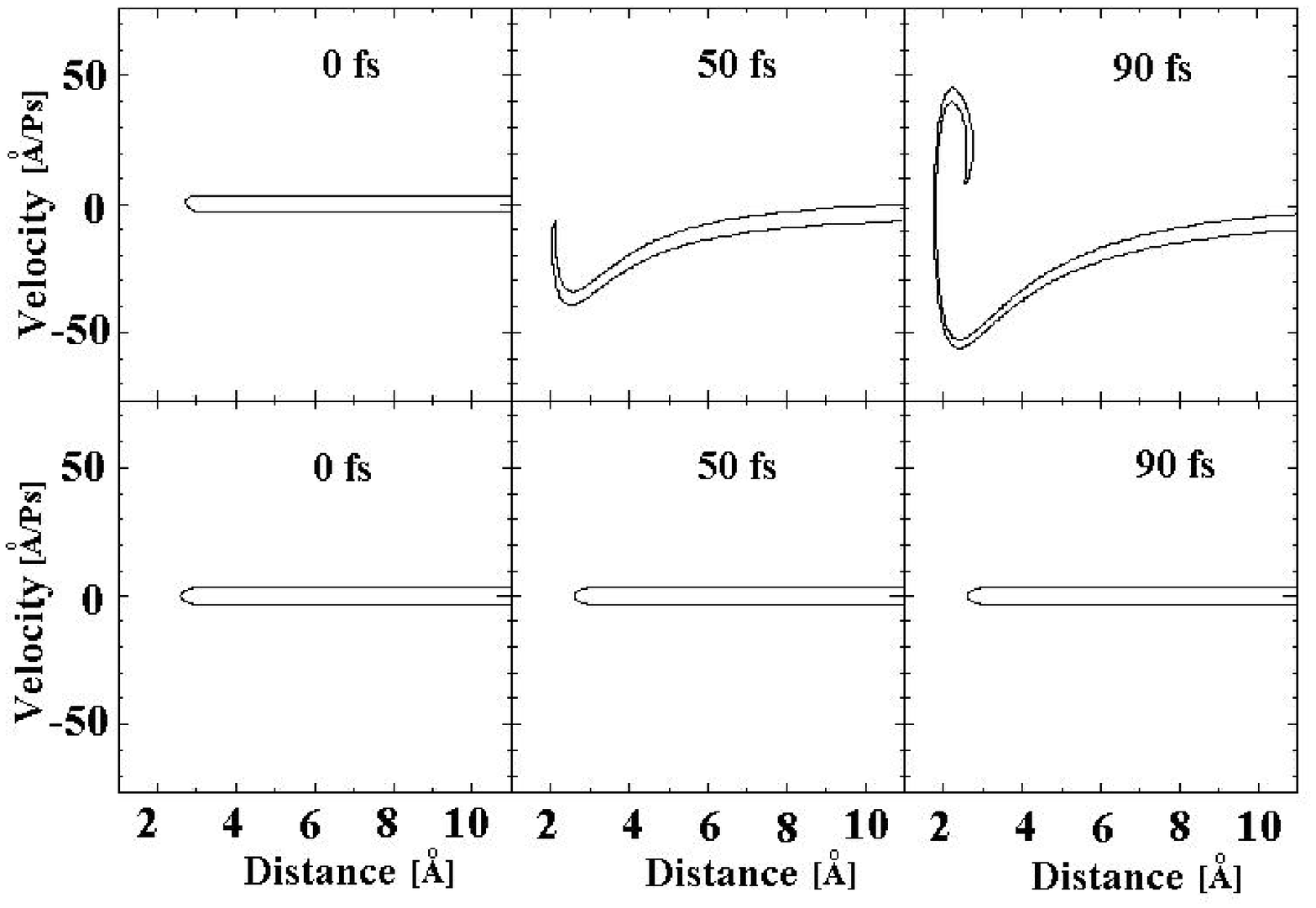}}
\end{figure}
The Wigner maps of the propagating X,B-states created by the pulse pair are
shown in Fig. 4. We take the inter-pulse delay equal to about a period ( 90
fs) of molecular vibration in the bottom of  B-state. The second impulse
transports the Wigner's B-state distribution partially back to continuum, as
it is, conserving its spiral shape.  The spiral branches are spreading on
the covalent X-curve and simultaneously squeezing in the momentum direction.
In addition, its optical field might induce a harpooning reaction by
producing a collisional pair of cation Kr and anion F. The Wigner
distribution on their B-binding term repeats the evolution following the
first impulse.
\begin{figure}
\caption{The Wigner distribution functions of KrF excimer created by the
impulse pair. The interpulse delay is equal to 90 fs i.e. about one
molecular period. The Wigner maps are drawn at 150 fs, 200 fs and 280 fs.
Upper panels present the harpooning reaction in the ionic B-states. Lower
panels show the spreading photofragments in covalent X-state.}
{\includegraphics[width=4.5in,height=3.5in]{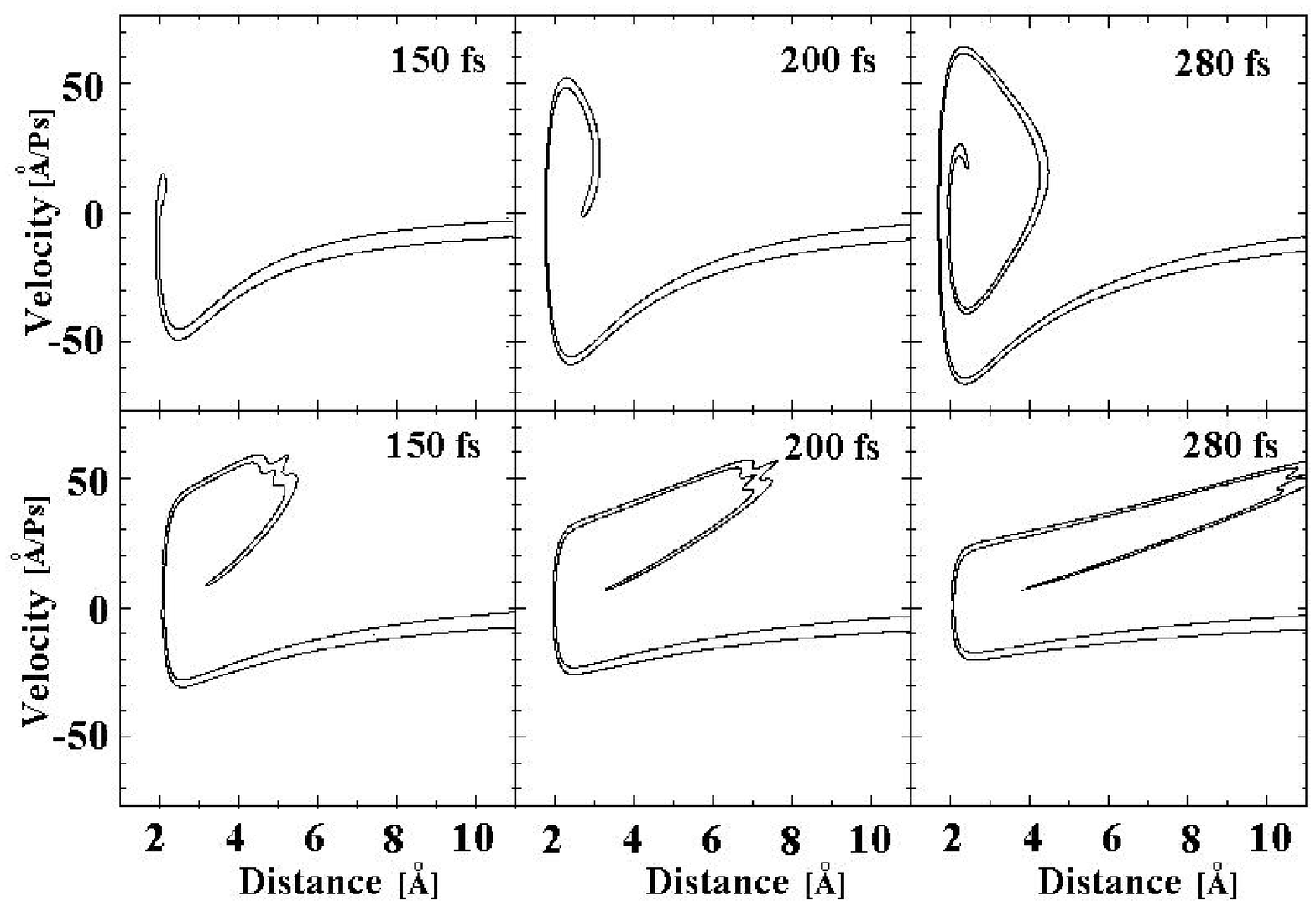}}
\end{figure}
In Fig. 5 we display the overlap of the molecular  densities in the phase
space as a function of time. The overlap of the Wigner's distributions of
the X and B -states quickly  tends to zero. The decay time after the first
impulse can be evaluated as a width at the half maximum height of the
integral Eq.(17). This time is about $20~fs$ long that conforms the estimate
based on the classical mechanics. Doing so, consider a particle moving with
a thermal velocity in the covalent X-state. Its wave packet is transported
to the binding B-term via a vertical Franck-Condon's transitions. Then, it
needs the time:
\begin{equation}
\tau_{decay}\approx\int_{R_{in}}^{R_{out}}\,dr/v_{b}(r)
\end{equation}
to destroy an alignment  of the Wigner's distributions while the wave packet
travels between the turning points in the quantum well. The turning points
$R_{in}\approx 2.15 \AA$ and  $R_{out}\approx 2.65 \AA$ are determined by
the thermal energy $E_{ther}=100 \,[cm^{-1}]$ of collision pairs of Kr and F
atoms. The wave packet loses its momentum correlation with nascent replica
of X state for time its journey between the turning points. Our first rough
estimate of about 40 fs is obtained by substituting to Eq.(19) the velocity:
$v_{b}(R)=(2(E_{ther}-V_{b}(R))/M)^{1/2}, $ and by taking the numerical
values of the reduced mass $M\approx15.5$ of the KrF molecule. Inset also
shows the rise and fall of the  overlaps of the Wigner's distributions after
the second ultrashort pulse. This occurs at twice time-delay, when the hook
map crosses the spiral branches of the X-state distribution. Our numerical
calculation indicates, that the photon echo signal is rather robust with
respect to a small deviations of the inter-pulse delay from the gating
conditions $T=T_{vib}$. This robustness can be attributed to anharmonicity
of the KrF(B) potential.
\begin{figure}
{\caption{The free radiation decay and photon echo of molecular
photoassociation of  KrF. The HWHM of $\mu^{(1)}$ polarization  is about 20
fs. The relevant $\mu^{(3)}$ polarization created by two $\delta$-like laser
fields having normalized pulse areas $\theta_{1,2}=1$ and inter-pulse delay
90 fs is shown in inset.}} {\includegraphics[width=4.in, height=3.0in
]{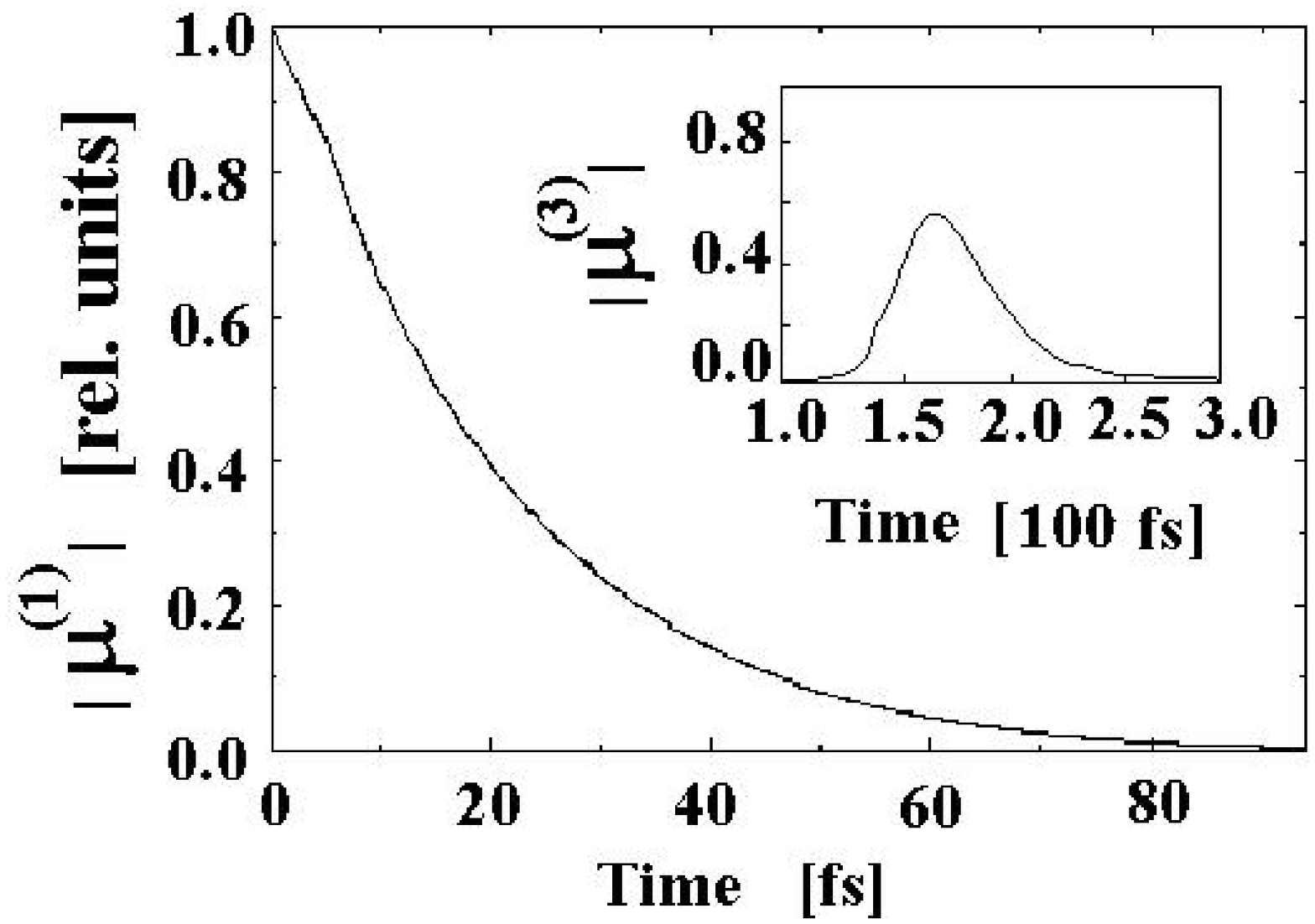}}
\end{figure}
\section{ Conclussions}
We have demonstrated how the ultrashort pulses can reverse the natural
tendency of the nonlinear optical transients to decay in photoassisted
half-collisions. The photon echo of molecular photoassociation presents a
paradigm of the coherent self-organization in optics. This occurs in the
continuum states initially residing in the Gibbs equilibrium ensemble. The
photoassosiation is followed by the delayed photodissociation giving rise
the correlation radius of propagating states to grow proportionally as the
inter-pulse time-delay. The latter results in a long lived transient
complex. Then, the echo signals manifests on the free-bound transitions
following the free radiation decay that is dictated by the resonant coupling
probability in accordance with the time dependent Fermi golden rule. This
delayed coherence is accounted for by the inverse predissociation amounting
to phenomena conceptually identical with the interference stabilization of
atoms (molecules) in the strong ionizing (dissociating) laser fields.

The femtosecond coherence of the laser assisted collisions is formed on the
molecular nano-scales and can be observed in optically thin samples. This
effect might provide a way to make a compact light compressor without
limitations imposed by the self-modulation effects in the optical
wave-guides. From the spectroscopic point of view, the photon echo of
photoassociation provides for unique data on the fly during the binary
half-collisions. The echo spikes splitting can exhibit an inhomogeneous
broadening inherent to the distribution of guest atoms in host solid with
the sub-wavelength resolution up to Angstrom's scale. This opens up a
possibility to trace the migration of open-shell atoms in rare gas matrices
and cage exit. Kindred phenomena might be studied by molecular dynamics
simulations $^{17}$ to provide a new avenue for the echo spectroscopy in
solids and liquids.

In brief, therein our major aim was not a quantitative agreement with an
experiment, but to broadly predict the general aspects of the photon echo of
photoassociation and to delineate those its features which depend on
inter-pulse delay of laser pumping, molecular potentials and temperature of
sample. The echo scheme has a merit to automate the wave packet preparation
making it a quite favorable for a practical design, where the pulse shaping
is not necessary. It is important to stress a geometrical aspect of the
effect, in which the directional vector ${\vec {k}_{1}-2\vec {k}_{2}}$ of
the echo coherence is distinguished from the incident wave vectors ${\vec{
k}_{1}}$ and ${\vec{k}_{2}}$ of the pump and delayed fields. This offers a
convenient experimental condition for observing spontaneous photons on dark
background with angular and temporal lag from the laser firing.  The details
of the signal may be varied from one molecular system to another, but the
broad conclusion is quite robust.
\begin{acknowledgments}
The work is supported in part by RFBR. AGR thanks Prof.
A.Giusti-Suzor and Dr. M.Macholm for the preprint \cite{3} and
useful discussions.
\end{acknowledgments}

\end{document}